\documentclass[12pt]{article}

\usepackage{amsmath,amsthm, amsfonts, amssymb, amsxtra,amsopn}
\usepackage{graphicx}
\usepackage{multirow}
\usepackage{algorithmic}
\usepackage{pgfplots}
\pgfplotsset{compat=1.12}
\usepackage{pgfplotstable}
\usepackage{adjustbox}

\PassOptionsToPackage{hyphens}{url}

\usepackage{hyperref}
\hypersetup{colorlinks=true,linkcolor=black,citecolor=black,urlcolor=blue,filecolor=black}
\hypersetup{pdfpagemode=UseNone,pdfstartview=}

\usepackage[tableposition=top]{caption}
\usepackage{enumitem}
\setlist[itemize]{noitemsep, topsep=0pt}

\advance\oddsidemargin by -0.25in
\advance\textwidth by 0.5in

\advance\topmargin by -0.25in
\advance\textheight by 0.5in

\def\z{\phantom{0}}

\def\z{\phantom{0}}
\def\O{{\cal O}}

\long\def\symbolfootnotetext[#1]#2{\begingroup%
\def\thefootnote{\fnsymbol{footnote}}\footnotetext[#1]{#2}\endgroup}

\title{Hidden Markov Models with Random Restarts vs Boosting for Malware Detection}

\author{Aditya Raghavan\footnotemark[1]\ \ \ 
Fabio Di Troia\footnotemark[1]\ \ \
Mark Stamp\footnotemark[1]\,\,\,\footnotemark[2]}

\begin{document}

\symbolfootnotetext[1]{Department of Computer Science, San Jose State University}
\symbolfootnotetext[2]{mark.stamp$@$sjsu.edu}

\maketitle

\abstract
Effective and efficient malware detection is at the forefront of research 
into building secure digital systems. As with many other fields, malware detection research has seen 
a dramatic increase in the application of machine learning algorithms. One machine learning 
technique that has been used widely in the field of pattern matching in general---and 
malware detection in particular---is hidden Markov models (HMMs). 
HMM training is based on a hill climb, 
and hence we can often improve a model 
by training multiple times with different initial values.
In this research, we compare boosted HMMs (using AdaBoost)
to HMMs trained with multiple random restarts,
in the context of malware detection. These techniques 
are applied to a variety of challenging malware datasets.
We find that random restarts perform surprisingly well
in comparison to boosting. Only in the most difficult
``cold start'' cases (where training data is severely limited)
does boosting appear to offer sufficient improvement to justify
its higher computational cost in the scoring phase.

\section{Introduction}\label{chap:introduction}

As of 2017, about~54\%\ of households worldwide had access to the Internet~\cite{ICU_Stats}. 
In terms of raw numbers, the count of Internet users has increased from around~1 billion in 2005 
to almost~3.6 billion in 2017~\cite{Statista_InternetUsers}. This trend of digitalization is sure 
to continue over the coming years, and soon virtually the entire world will be connected to the Internet. 

The proliferation of computers and the widespread 
use of the Internet have resulted in the digitalization of many services. 
Business applications are obvious, but highly digitized services also
include essentials such as the power grid, dams, traffic lights,
and so on. The Internet of Things (IoT) 
promises to connect nearly every aspect of life 
to the Internet---as of~2018, there 
are about~23 billion IoT connected devices~\cite{Statista_IoTDevices}. 
This reliance on digitalization and automation brings with it a set of challenges,
and chief among these challenges is digital security.

Today, bad actors can exploit our reliance on technology for financial gain, and
there are credible predictions of cyber-warfare being 
a leading mode of attack in future conflicts~\cite{Rand_Cyber_Warfare}. 
Malicious software, or malware, is the driving force behind a vast array of digital security issues. 

According to~\cite{MalwarePandaStats}, 
one in three computers worldwide is affected 
by malware. While the cost of malware is notoriously difficult to quantify, estimates are
that financial losses due to cyber crime will reach a staggering~\$6 trillion annually by 
the year~2021~\cite{AccentureCyberCrime}. Not surprisingly, 
spending on cyber defenses is also expected to increase---it is  
estimated that such expenditures will exceed~\$1 trillion in~2021~\cite{CyberSecExpenses}. 

A wide variety of different types of malware affect computing systems.
Such malware includes adware, spyware, Trojans, viruses, worms, 
ransomware, and many others~\cite{Malware101}.
%
Antivirus software, firewalls, and intrusion detection systems are used in attempts 
to keep systems secure. Antivirus software generally relies primarily on signature detection 
(i.e., pattern matching) to detect malware. However, there are many advanced forms of malware 
that can evade signature-based detection~\cite{AycockBook}.

Machine learning techniques can be used to improve on signature-based detection~\cite{HMMDetection}. 
Hidden Markov models (HMMs) are one popular machine learning technique that has
been successfully applied to the malware detection 
problem~\cite{AnnachhatreAS15,HMMDetection,survey,IntroMLInfoSec},
as well as a wide variety of other information security 
problems~\cite{Ariu2011,BF2009,Cho2003,Hu2009,okamoto2007framework,Okamoto2011,Posadas2006,
RanaS14,SimovaSP05,Sperotto2009,Srivastava2008}.
In this research, we consider the effectiveness of 
malware detection based on HMMs with multiple random restarts. We compare this 
random restarts approach to combining HMMs using the well-known 
AdaBoost algorithm~\cite{InfoSecAdaBoost}. Interestingly, 
it appears that boosted HMMs have not previously received much attention in the
information security domain~\cite{BoostedHMMIDS}. 

We consider a variety of experiments to compare multiple random restarts
to boosted HMMs. Our experiments include the so-called ``cold start'' problem, 
where limited training data is available. We believe that all of our experiments provide 
realistic and challenging test cases for comparing the techniques under consideration.

The remainder of this paper is organized as follows. In Section~\ref{chap:background},
we provide relevant background information,
including brief introductions to hidden Markov models and AdaBoost.
Section~\ref{chap:exp_results} describes our 
experimental setup and results. In this section, we also
provide some discussion and analysis of our results. 
Finally, in Section~\ref{chap:cl_fw}, we present our conclusion and mention 
possible future work.

\section{Background}\label{chap:background}

Machine learning can be viewed as a form of statistical discrimination where a ``machine'' or algorithm
does the hard work, rather than a human analyst~\cite{IntroMLInfoSec}.
Complex problems such as character recognition and voice identification can be effectively
solved using various machine learning techniques~\cite{SpeechRecogPhoneticMarkov, CharRecogUsingHMM}. 
In the field of information security, machine learning has become a fundamental tool in many areas  
of research, including malware detection and analysis, and intrusion 
detection~\cite{HMMDetection,BoostedHMMIDS,AdaBoostIDS}.

Next, we briefly introduce hidden Markov models and AdaBoost. These are both
popular machine learning techniques that have found widespread application to problems
in information security.

\subsection{Hidden Markov Model}\label{sec:hmm}

\begin{figure}[!htb]
  \begin{center}
    \begin{tikzpicture}[scale=1.0]
    
    \draw[thick,color=blue] (0,0) rectangle (1,1);
    \draw[thick,color=blue] (2.5,0) rectangle (3.5,1);
    \draw[thick,color=blue] (5,0) rectangle (6,1);
    \draw[thick,color=blue] (10,0) rectangle (11,1);

    \draw[thick,color=green] (0.5,4.5) circle (0.575);
    \draw[thick,color=green] (3,4.5) circle (0.575);
    \draw[thick,color=green] (5.5,4.5) circle (0.575);
    \draw[thick,color=green] (10.5,4.5) circle (0.575);
    
    \node at (0.5,0.5){$\O_0$};
    \node at (3,0.5){$\O_1$};
    \node at (5.5,0.5){$\O_2$};
    \node at (8,0.5){$\cdots$};
    \node at (10.5,0.5){$\O_{T-1}$};

    \node at (0.5,4.5){$X_0$};
    \node at (3,4.5){$X_1$};
    \node at (5.5,4.5){$X_2$};
    \node at (8,4.5){$\cdots$};
    \node at (10.5,4.5){$X_{T-1}$};
       
    \node at (1.7,4.8){$A$};
    \node at (4.2,4.8){$A$};
    \node at (6.7,4.8){$A$};
    \node at (9.2,4.8){$A$};
    
    \node at (0.2,2.1){$B$};
    \node at (2.7,2.1){$B$};
    \node at (5.2,2.1){$B$};
    \node at (10.2,2.1){$B$};
    
     \draw[thick,color=black,->] (1.075,4.5) -- (2.425,4.5);
     \draw[thick,color=black,->] (3.575,4.5) -- (4.925,4.5);
     \draw[thick,color=black,->] (6.075,4.5) -- (7.425,4.5);
     \draw[thick,color=black,->] (8.575,4.5) -- (9.925,4.5);

     \draw[thick,color=black,->] (0.5,3.925) -- (0.5,1);
     \draw[thick,color=black,->] (3.0,3.925) -- (3.0,1);
     \draw[thick,color=black,->] (5.5,3.925) -- (5.5,1);
     \draw[thick,color=black,->] (10.5,3.925) -- (10.5,1);

    \draw[thick,dashed,color=red] (-0.3,3) -- (11.2,3);
   
    \end{tikzpicture}
  \end{center}
  \caption{Hidden Markov model}\label{fig:hidden_Markov_model}
\end{figure}
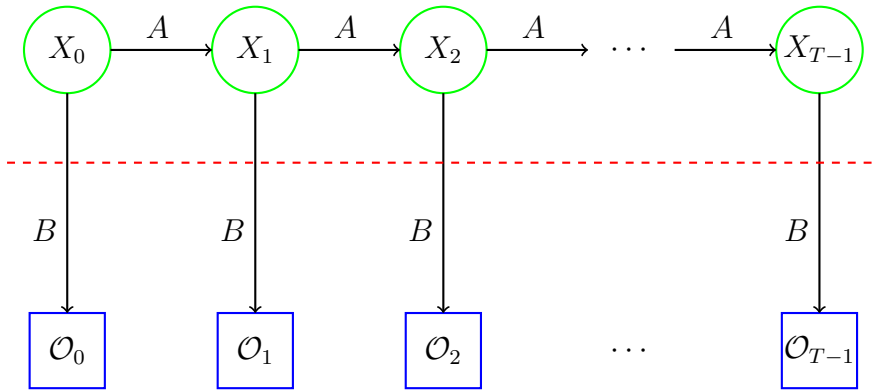

In a Markov process of order one, the current state depends only on
the previous state, and the state transition probabilities are based on
fixed, discrete probability distributions. 
As the name suggests, in a hidden Markov model, we cannot
directly observe the state sequence, but we do have access to 
a series of observations that are related to the hidden states
via discrete probability distributions.
A generic view of an HMM is given in Figure~\ref{fig:hidden_Markov_model},
with the relevant notation defined in Table~\ref{tab:hidden_Markov_model_notations}.

\begin{table}[!htb]
  \caption{HMM notation}\label{tab:hidden_Markov_model_notations}
  \centering
  \begin{tabular}{cl} \hline\hline
    Notation & \hspace*{0.5in}Description\\ \hline
    $T$ & Length of the observation sequence\\
    $N$ & Number of states in the model\\
    $M$ & Number of observation symbols\\
    $Q$ & Distinct states of the Markov process, $q_0,q_1,\ldots,q_{N-1}$\\
    $V$ & Possible observations, assumed to be $0,1,\ldots,M-1$\\
    $A$ & State transition probabilities\\
    $B$ & Observation probability matrix\\
    $\pi$ & Initial state distribution\\
    $\O$ & Observation sequence, $\O_0,\O_1,\ldots,\O_{T-1}$ \\ \hline\hline
  \end{tabular}
\end{table}

Using the notation in Table~\ref{tab:hidden_Markov_model_notations}, 
the state transition matrix~$A$ is of size~$N\times N$, 
the observation probability matrix~$B$ is~$N\times M$,
and the initial distribution matrix~$\pi$ is~$1\times N$.
These matrices are all row stochastic, and they define an HMM;
thus, we denote an HMM as~$\lambda=(A,B,\pi)$.

The HMM training process is a hill climb, and we typically initialize the
elements of~$A$, $B$, and~$\pi$ to approximately uniform, that is,
$a_{i,j}\approx 1/N$, $b_{i,j}\approx 1/M$ and~$\pi_{1,j}\approx 1/N$,
with the row stochastic requirement enforced.
Since HMM training is a hill climb, we might obtain better
results by training multiple models with different initial values,
and selecting the best of the resulting models. Such an 
``HMM with random restarts'' approach has been used,
for example, in the analysis of classic 
substitution ciphers~\cite{BKK2013,VobbilisettyTLV17}.
However, as far as the authors are aware, such an
approach has not been explicitly applied in the field of 
malware detection.

Previous research employing hidden Markov models includes
a wide array of pattern matching problems. For example,
HMMs have been used to distinguish handwritten letters
with high accuracy~\cite{CharRecogUsingHMM}.
Voice recognition is another area where HMMs 
are a reliable and strong performer~\cite{HMMSpeechRecogAlgo}.
And, as previously noted, HMMs are often
used in malware research.

In this paper, we consider the malware detection problem and analyze
the effectiveness of generating~1000 HMMs with random
restarts. We compare this random restarts approach to the 
average case, where a single HMM is trained.
We also consider the case where the~1000
HMMs are combined using AdaBoost.

Next, we briefly introduce the key ideas behind the AdaBoost algorithm.
While AdaBoost is often used with decision trees, the technique
will work with any type of classifiers---in this paper, 
the boosted classifiers are based on HMMs.


\subsection{AdaBoost}\label{sec:adaboost}

Boosting is the process of combining multiple (weak) classifiers to obtain a stronger classifier.
Any classifier that performs better than a coin flip can be used, and if a sufficient number of
such classifiers are available, adaptive boosting, or AdaBoost, can generate an arbitrarily strong 
classifier~\cite{InfoSecAdaBoost}. There are other boosting techniques, including extreme gradient
boosting (XGBoost), but AdaBoost is certainly the best known boosting technique.

At each iteration, AdaBoost selects the 
``best'' classifier from those available (i.e., unused), where
``best'' is defined as the classifier that most improves on the overall accuracy
of the new, combined classifier. That is, AdaBoost greedily
selects a classifier that does the most to improve on the current
iteration of the constructed classifier. In AdaBoost the selected classifiers are 
combined as a weighted linear combination, where an optimal weight is
calculated at each iteration, with all previously computed weights fixed.
While AdaBoost has many desirable properties, one inherent problem is
that errors in the training data tend to grow, due to the iterative
nature of the algorithm.

%

\begin{figure}[!htbp]
	\centering
    \begin{tikzpicture}[scale=0.6]
    \begin{axis}[width=1.0\textwidth,height=1.0\textwidth,
    		      xmin=1.0,xmax=200.0,
                       ymin=50.0,ymax=102.0, legend pos=south east,
                       x tick label style={
    			/pgf/number format/.cd,
			/pgf/number format/1000 sep={},
    			fixed,
    			fixed zerofill,
    			precision=0},                       
                       xlabel={Iteration},ylabel={Correct Classifications},
                       y tick label style={
    			/pgf/number format/.cd,
    			fixed,
    			fixed zerofill,
    			precision=0}] 
\addplot[color=blue,ultra thick,
	mark=none] coordinates {
(1, 52)
(2, 53)
(3, 56)
(4, 56)
(5, 57)
(6, 61)
(7, 55)
(8, 58)
(9, 59)
(10, 61)
(11, 61)
(12, 63)
(13, 60)
(14, 59)
(15, 64)
(16, 64)
(17, 66)
(18, 65)
(19, 68)
(20, 66)
(21, 68)
(22, 70)
(23, 70)
(24, 64)
(25, 69)
(26, 70)
(27, 68)
(28, 70)
(29, 70)
(30, 71)
(31, 72)
(32, 71)
(33, 71)
(34, 73)
(35, 76)
(36, 79)
(37, 76)
(38, 78)
(39, 80)
(40, 81)
(41, 81)
(42, 83)
(43, 78)
(44, 84)
(45, 81)
(46, 85)
(47, 81)
(48, 86)
(49, 84)
(50, 85)
(51, 83)
(52, 85)
(53, 87)
(54, 86)
(55, 88)
(56, 88)
(57, 86)
(58, 88)
(59, 88)
(60, 87)
(61, 87)
(62, 88)
(63, 88)
(64, 89)
(65, 91)
(66, 93)
(67, 92)
(68, 90)
(69, 91)
(70, 89)
(71, 92)
(72, 92)
(73, 94)
(74, 92)
(75, 95)
(76, 95)
(77, 94)
(78, 94)
(79, 95)
(80, 94)
(81, 96)
(82, 94)
(83, 98)
(84, 99)
(85, 97)
(86, 97)
(87, 100)
(88, 97)
(89, 99)
(90, 99)
(91, 100)
(92, 100)
(93, 98)
(94, 100)
(95, 99)
(96, 98)
(97, 98)
(98, 100)
(99, 100)
(100, 99)
(101, 100)
(102, 99)
(103, 100)
(104, 99)
(105, 99)
(106, 99)
(107, 99)
(108, 98)
(109, 99)
(110, 99)
(111, 99)
(112, 99)
(113, 99)
(114, 98)
(115, 98)
(116, 98)
(117, 99)
(118, 98)
(119, 99)
(120, 99)
(121, 99)
(122, 100)
(123, 99)
(124, 99)
(125, 99)
(126, 100)
(127, 99)
(128, 100)
(129, 99)
(130, 100)
(131, 99)
(132, 100)
(133, 100)
(134, 100)
(135, 100)
(136, 100)
(137, 100)
(138, 100)
(139, 100)
(140, 100)
(141, 100)
(142, 100)
(143, 100)
(144, 100)
(145, 100)
(146, 100)
(147, 100)
(148, 100)
(149, 100)
(150, 100)
(151, 100)
(152, 100)
(153, 100)
(154, 100)
(155, 100)
(156, 100)
(157, 100)
(158, 99)
(159, 99)
(160, 99)
(161, 100)
(162, 100)
(163, 100)
(164, 100)
(165, 100)
(166, 100)
(167, 100)
(168, 100)
(169, 100)
(170, 100)
(171, 100)
(172, 100)
(173, 100)
(174, 100)
(175, 100)
(176, 100)
(177, 100)
(178, 100)
(179, 100)
(180, 100)
(181, 100)
(182, 100)
(183, 100)
(184, 100)
(185, 100)
(186, 100)
(187, 100)
(188, 100)
(189, 100)
(190, 100)
(191, 100)
(192, 100)
(193, 100)
(194, 100)
(195, 100)
(196, 100)
(197, 100)
(198, 100)
(199, 100)
};
\addlegendentry{$L = 1000$}
\addplot[color=green,ultra thick,
	mark=none] coordinates {
(1, 52)
(2, 50)
(3, 55)
(4, 53)
(5, 52)
(6, 55)
(7, 55)
(8, 61)
(9, 57)
(10, 63)
(11, 56)
(12, 65)
(13, 65)
(14, 63)
(15, 57)
(16, 67)
(17, 61)
(18, 69)
(19, 67)
(20, 67)
(21, 67)
(22, 65)
(23, 62)
(24, 72)
(25, 69)
(26, 72)
(27, 70)
(28, 71)
(29, 70)
(30, 74)
(31, 72)
(32, 71)
(33, 73)
(34, 74)
(35, 74)
(36, 77)
(37, 73)
(38, 79)
(39, 80)
(40, 79)
(41, 75)
(42, 79)
(43, 79)
(44, 79)
(45, 77)
(46, 78)
(47, 76)
(48, 82)
(49, 84)
(50, 83)
(51, 79)
(52, 83)
(53, 81)
(54, 81)
(55, 81)
(56, 82)
(57, 87)
(58, 83)
(59, 83)
(60, 81)
(61, 84)
(62, 83)
(63, 88)
(64, 87)
(65, 89)
(66, 87)
(67, 87)
(68, 87)
(69, 84)
(70, 87)
(71, 88)
(72, 88)
(73, 89)
(74, 88)
(75, 89)
(76, 90)
(77, 88)
(78, 88)
(79, 89)
(80, 90)
(81, 91)
(82, 89)
(83, 88)
(84, 91)
(85, 86)
(86, 89)
(87, 92)
(88, 92)
(89, 91)
(90, 93)
(91, 92)
(92, 94)
(93, 94)
(94, 91)
(95, 92)
(96, 94)
(97, 93)
(98, 91)
(99, 94)
(100, 91)
(101, 94)
(102, 96)
(103, 95)
(104, 96)
(105, 96)
(106, 97)
(107, 97)
(108, 95)
(109, 97)
(110, 96)
(111, 97)
(112, 98)
(113, 97)
(114, 97)
(115, 96)
(116, 98)
(117, 94)
(118, 97)
(119, 95)
(120, 96)
(121, 97)
(122, 97)
(123, 98)
(124, 98)
(125, 97)
(126, 98)
(127, 99)
(128, 98)
(129, 98)
(130, 97)
(131, 98)
(132, 96)
(133, 99)
(134, 97)
(135, 98)
(136, 98)
(137, 99)
(138, 99)
(139, 98)
(140, 100)
(141, 100)
(142, 99)
(143, 98)
(144, 100)
(145, 99)
(146, 100)
(147, 98)
(148, 99)
(149, 98)
(150, 99)
(151, 100)
(152, 100)
(153, 100)
(154, 100)
(155, 99)
(156, 99)
(157, 98)
(158, 99)
(159, 98)
(160, 99)
(161, 100)
(162, 100)
(163, 100)
(164, 100)
(165, 100)
(166, 100)
(167, 100)
(168, 100)
(169, 100)
(170, 100)
(171, 100)
(172, 100)
(173, 100)
(174, 99)
(175, 99)
(176, 99)
(177, 100)
(178, 100)
(179, 100)
(180, 100)
(181, 100)
(182, 100)
(183, 99)
(184, 100)
(185, 100)
(186, 100)
(187, 100)
(188, 100)
(189, 100)
(190, 100)
(191, 100)
(192, 100)
(193, 100)
(194, 99)
(195, 100)
(196, 99)
(197, 100)
(198, 100)
(199, 100)
};
\addlegendentry{$L = 500$}
\addplot[color=red,ultra thick,
	mark=none] coordinates {
(1, 52)
(2, 54)
(3, 52)
(4, 56)
(5, 56)
(6, 56)
(7, 61)
(8, 56)
(9, 60)
(10, 62)
(11, 63)
(12, 63)
(13, 60)
(14, 61)
(15, 63)
(16, 62)
(17, 62)
(18, 60)
(19, 65)
(20, 63)
(21, 60)
(22, 64)
(23, 66)
(24, 66)
(25, 68)
(26, 66)
(27, 62)
(28, 67)
(29, 68)
(30, 69)
(31, 70)
(32, 70)
(33, 72)
(34, 68)
(35, 70)
(36, 72)
(37, 70)
(38, 71)
(39, 72)
(40, 69)
(41, 70)
(42, 71)
(43, 75)
(44, 69)
(45, 78)
(46, 74)
(47, 77)
(48, 73)
(49, 71)
(50, 72)
(51, 74)
(52, 73)
(53, 76)
(54, 73)
(55, 74)
(56, 75)
(57, 76)
(58, 79)
(59, 81)
(60, 78)
(61, 81)
(62, 80)
(63, 85)
(64, 79)
(65, 80)
(66, 82)
(67, 77)
(68, 83)
(69, 79)
(70, 79)
(71, 77)
(72, 80)
(73, 79)
(74, 82)
(75, 78)
(76, 82)
(77, 83)
(78, 84)
(79, 82)
(80, 83)
(81, 84)
(82, 84)
(83, 85)
(84, 85)
(85, 82)
(86, 83)
(87, 79)
(88, 85)
(89, 85)
(90, 87)
(91, 87)
(92, 85)
(93, 86)
(94, 87)
(95, 85)
(96, 85)
(97, 84)
(98, 85)
(99, 84)
(100, 86)
(101, 85)
(102, 83)
(103, 82)
(104, 83)
(105, 86)
(106, 87)
(107, 85)
(108, 84)
(109, 85)
(110, 87)
(111, 86)
(112, 87)
(113, 84)
(114, 86)
(115, 84)
(116, 88)
(117, 85)
(118, 87)
(119, 79)
(120, 86)
(121, 83)
(122, 86)
(123, 85)
(124, 88)
(125, 86)
(126, 85)
(127, 86)
(128, 86)
(129, 86)
(130, 87)
(131, 86)
(132, 88)
(133, 88)
(134, 88)
(135, 86)
(136, 89)
(137, 88)
(138, 87)
(139, 84)
(140, 85)
(141, 86)
(142, 86)
(143, 87)
(144, 88)
(145, 87)
(146, 86)
(147, 87)
(148, 86)
(149, 84)
(150, 86)
(151, 87)
(152, 86)
(153, 84)
(154, 84)
(155, 87)
(156, 83)
(157, 85)
(158, 85)
(159, 86)
(160, 86)
(161, 86)
(162, 87)
(163, 89)
(164, 87)
(165, 91)
(166, 89)
(167, 89)
(168, 87)
(169, 86)
(170, 86)
(171, 85)
(172, 84)
(173, 85)
(174, 86)
(175, 85)
(176, 87)
(177, 86)
(178, 85)
(179, 84)
(180, 83)
(181, 84)
(182, 83)
(183, 84)
(184, 82)
(185, 84)
(186, 82)
(187, 83)
(188, 81)
(189, 81)
(190, 81)
(191, 80)
(192, 81)
(193, 80)
(194, 80)
(195, 80)
(196, 80)
(197, 80)
(198, 80)
(199, 80)
};
\addlegendentry{$L = 250$}
    \end{axis}
    \end{tikzpicture}
\caption{Correct classifications vs iteration~\cite{InfoSecAdaBoost}}\label{fig:ccIter1001000}	
\end{figure}
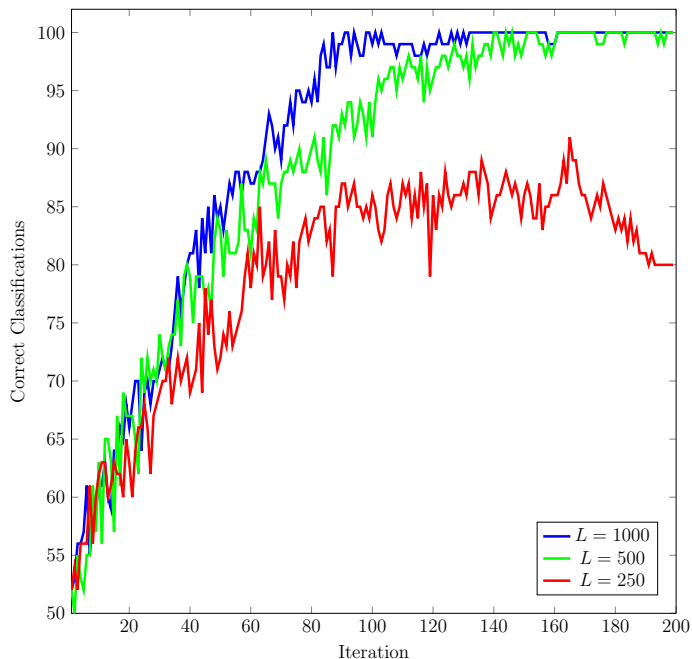

Although AdaBoost is a greedy algorithm, it is worth noting that it is not a hill climb.
Hence, at any given iteration, it is possible that the resulting classifier will be worse
than at the previous iteration. Figure~\ref{fig:ccIter1001000} shows the accuracy of
AdaBoost as a function of the iteration number in three different cases.
In each case, the same set of~$n=100$ labeled samples was used,  
with the number of (extremely weak) classifiers being~$L=250$, 
$L=500$, and~$L=1000$ for the three different cases, which appear as
the red, green, and blue graphs in Figure~\ref{fig:ccIter1001000}, respectively.
The dips in the accuracy, can be fairly substantial,
and if we do not have a sufficient number of classifiers available,
the results will suffer, as can be seen by the~$L=250$ case in Figure~\ref{fig:ccIter1001000}. 
The tutorial~\cite{InfoSecAdaBoost} discusses
these and related issues in more detail.

There have been many successful applications of boosting algorithms. 
One such example uses AdaBoost to improve the selection of 
features in a vision based application~\cite{AdaBoostFeatureSelection}. 
Security-related applications of AdaBoost can be found in~\cite{BoostedHMMIDS,AdaBoostIDS},
where several classifiers based on Gaussian mixture models are combined 
into a stronger classifier for network intrusion detection.

\subsection{Evaluation Criteria}

We use accuracy as an evaluation metric for some of the experimental
results that we present in Section~\ref{chap:exp_results}.
For an experiment on a labeled dataset,
$$
\mbox{accuracy} = \dfrac{\mbox{TP} + \mbox{TN}}{\mbox{TP} + \mbox{TN} + \mbox{FP} + \mbox{FN}}
$$
where
\begin{align*}
\mbox{TP} =& \mbox{true positives}, \mbox{TN} = \mbox{true negatives} \\
\mbox{FP} =& \mbox{false positives}, \mbox{FN} = \mbox{false negatives}
\end{align*}
Accuracy is an intuitive measure, as it is simply
the ratio of correct classifications to the total number of classifications.

We employ receiver operating characteristic (ROC) curve analysis in
all of our experiments. An
ROC curve is obtained from a scatterplot by graphing the true positive rate (TPR) 
versus the false positive rate (FPR) as the threshold varies through the range of values.
These rates are computed as
$$
  \mbox{TPR} = \frac{\mbox{TP}}{\mbox{TP} + \mbox{FN}}
  \mbox{\ \  and\ \ }
  \mbox{FPR} = \frac{\mbox{FP}}{\mbox{FP} + \mbox{TN}}
$$

The area under the ROC curve (AUC) ranges from~0 to~1, with~1 indicating ideal separation, that is,
a threshold exists for which no misclassifications occur. An AUC of~0.5 indicates that the underlying 
binary classifier is no better than flipping a coin. Also, note that an AUC of~$x < 0.5$ will yield
an AUC of~$1-x > 0.5$ if we simply reverse the sense of the binary classifier. 
The AUC can be interpreted as the probability that a randomly selected positive instance
scores higher than a randomly selected negative instance~\cite{auc}.


\section{Experiments and Results}\label{chap:exp_results}

In this section, we give detailed experimental results comparing HMMs
with multiple random restarts to boosted HMMs. First, we discuss the
basic parameters of the experiments; then, we present three sets of 
experiments.

\subsection{Dataset and Features}

All of the experiments here are based on malware samples 
from the Malicia dataset~\cite{maliciaDataset},
along with a representative collection of benign samples.
The benign samples consist of Windows system~32 executables
collected from a fresh install,
while the malicious families are the following.
\begin{description}
\item[Cridex] is a Trojan that creates a backdoor and collects sensitive
information, such as details related to online banking. The resulting
information can then be transmitted to a third party~\cite{cridex}.
\item[Harebot] is a backdoor that can yield remote access to an infected system. 
Due to its large number of features, Harebot is also 
sometimes considered to be a rootkit~\cite{harebot}. 
\item[Security Shield] is a spyware Trojan that claims to be anti-virus software
and reports fake virus detection results to the user. Security Shield attempts to
coerce the user into purchasing software~\cite{securityshield}.
\item[Zbot] is a Trojan horse that compromises a system 
by downloading configuration files or updates. Zbot, which is also known as Zeus, 
is stealthy malware that attempts to hide in the file system~\cite{zbot}. 
\item[ZeroAccess] is a Trojan horse that makes use of an advanced rootkit to hide its
presence. ZeroAccess can create a new (hidden) file system,
install a backdoor, 
and download additional malware, among other features~\cite{zeroaccess}.
\end{description}

Table~\ref{tab:maliciaTable} lists the number of samples from
each malware family considered, as well as the number of 
benign samples. Note that the number of benign samples is larger than the
number of malware samples for three of the five malware families under consideration.
All of our subsequent experiments and analysis are conducted on a per-family basis.

\begin{table}[htb]
	\caption{Dataset \label{tab:maliciaTable}}
	\centering
		\begin{tabular}{llc}\hline\hline
			Family & Type & Samples\\ \hline
			Cridex & Trojan & \z\z74 \\
			Harebot & Backdoor & \z\z53 \\
			Security Shield & Spyware & \z\z58 \\
			Zbot & Trojan & 2316 \\
			ZeroAccess & Trojan & 1305 \\ \hline
			Benign & \ \ \ --- & \z107 \\ \hline\hline
		\end{tabular}
\end{table}

Good detection results can be obtained for some of the larger families (e.g., ZeroAccess)
in the Malicia dataset, but the smaller families (e.g., Cridex and Harebot) have been shown to be
challenging~\cite{Naman_masters,Naman,Deebiga_masters,Deebiga}.
In addition, HMMs have generally been found to be competitive with 
many other proposed malware detection 
techniques~\cite{BaysaLS13,KalbhorAFJS15,ShanmugamLS13,SinghTVAS16}.
Our initial experimental results, which appear below in Section~\ref{sec:initial_exp},
are consistent with this previous work. 
However, the main contribution of this paper is to be found in the relative differences 
between boosting and random restarts in the most challenging test cases,
which are discussed in Sections~\ref{sec:morphing} and~\ref{sec:cold_start_problem}.

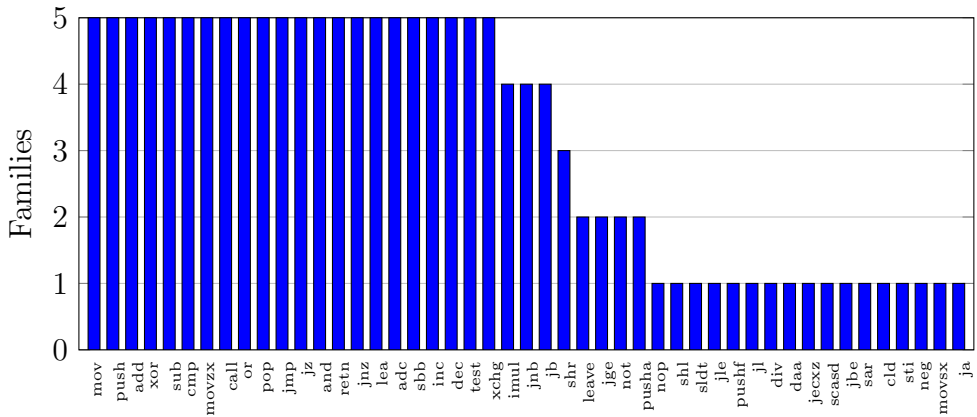
\begin{figure}[htbp]
\centering
\begin{tikzpicture}
    \begin{axis}[
        width  = 0.9*\textwidth,
        height = 6cm,
        major x tick style = transparent,
        ybar=4*\pgflinewidth,
        bar width=4.5pt,
        ymajorgrids = true,
        ylabel = {Families},
        symbolic x coords={
mov,
push,
add,
xor,
sub,
cmp,
movzx,
call,
or,
pop,
jmp,
jz,
and,
retn,
jnz,
lea,
adc,
sbb,
inc,
dec,
test,
xchg,
imul,
jnb,
jb,
shr,
leave,
jge,
not,
pusha,
nop,
shl,
sldt,
jle,
pushf,
jl,
div,
daa,
jecxz,
scasd,
jbe,
sar,
cld,
sti,
neg,
movsx,
ja
},
	y tick label style={
    	/pgf/number format/.cd,
   	fixed,
   	fixed zerofill,
    	precision=0},
        xtick = data,
        ytick = {0,1,2,3,4,5},
        x tick label style={rotate=90,anchor=north east, inner sep=0mm,font=\tiny},
        scaled y ticks = false,
        enlarge x limits=0.0175,
        ymin=0,
        ymax=5.0,
    ]
        \addplot[fill=blue]
            coordinates {
(mov,5)
(push,5)
(add,5)
(xor,5)
(sub,5)
(cmp,5)
(movzx,5)
(call,5)
(or,5)
(pop,5)
(jmp,5)
(jz,5)
(and,5)
(retn,5)
(jnz,5)
(lea,5)
(adc,5)
(sbb,5)
(inc,5)
(dec,5)
(test,5)
(xchg,5)
(imul,4)
(jnb,4)
(jb,4)
(shr,3)
(leave,2)
(jge,2)
(not,2)
(pusha,2)
(nop,1)
(shl,1)
(sldt,1)
(jle,1)
(pushf,1)
(jl,1)
(div,1)
(daa,1)
(jecxz,1)
(scasd,1)
(jbe,1)
(sar,1)
(cld,1)
(sti,1)
(neg,1)
(movsx,1)
(ja,1)
};
\end{axis}
\end{tikzpicture}
\caption{Number of families for each of top 30 opcodes\label{bar:top_30}} 
\end{figure}

The feature used in all of our experiments is the mnemonic opcode sequence.
Following previous work, for each family, we use the top~30 most common opcodes, 
and group all remaining
opcodes together as ``other''~\cite{ShanmugamLS13}. Note that this 
will generally give us a different set of opcodes for each family, but the overlap 
between families is significant. For a typical set of
experiments, the distinct top~30 opcodes and number of families that
each appears in is given in Figure~\ref{bar:top_30}.
In this case, we see that~22 of the top~30 opcodes
are common to all five malware families.

Executables are disassembled and mnemonic opcodes are 
extracted. Then the top~30 opcodes are determines, and
any opcodes outside of the top~30 are replaced with ``other.''
The percentage of opcodes in each family 
that are found among the top~30 are given in Table~\ref{tab:BaseExpOpcodes}.
We see that in each case, the vast majority of opcodes lie within
the top~30.

\begin{table}[htb]
	\caption{Top opcodes frequency\label{tab:BaseExpOpcodes}}
	\centering
		\begin{tabular}{lc}\hline\hline
			\multirow{2}{*}{Family} & \multicolumn{1}{c}{Top 30 opcodes}\\ 
			                         & \multicolumn{1}{c}{(percentage)}\\ \hline
			Cridex & 95\\
			Harebot & 94 \\
			Security Shield & 96 \\
			Zbot & 94 \\
			ZeroAccess & 96 \\ \hline\hline
		\end{tabular}
\end{table}

For all of the experiments reported below, we use 5-fold cross validation. 
Cross validation serves to smooth any bias in the data, and also provides 
us with the maximum possible number of independent test cases~\cite{IntroMLInfoSec}.

Again, all HMMs are trained on extracted opcode sequences. To score a given
sample against a specific HMM, we extract the opcode sequence from the sample
under analysis, and score the resulting sequence against the model, 
then normalize the score by the length of the
opcode sequence. This gives us a log-likelihood per opcode (LLPO) score.
Since an HMM score is length dependent, the LLPO score enables us to
directly compare samples with differing numbers of opcodes.

\subsection{Initial Experiments}\label{sec:initial_exp}

\begin{figure}[htbp]
\centering
\begin{tikzpicture}
    \begin{axis}[
        width  = 0.65*\textwidth,
        height = 8.5cm,
        major x tick style = transparent,
        ybar=4*\pgflinewidth,
        bar width=8pt,
        ymajorgrids = true,
        ylabel = {AUC},
        symbolic x coords={Cridex, Harebot, Security Shield, Zbot, ZeroAccess},
	y tick label style={
    	/pgf/number format/.cd,
   	fixed,
   	fixed zerofill,
    	precision=2},
        xtick = data,
        x tick label style={rotate=45,anchor=north east, inner sep=0mm,font=\footnotesize},
        scaled y ticks = false,
        enlarge x limits=0.135,
        ymin=0,
        ymax=1.05,
        legend cell align=left,
        legend pos=south east,
    ]
        \addplot[fill=red]
            coordinates {
(Cridex,0.544054)
(Harebot,0.6042724)
(Security Shield,0.6134088)
(Zbot,0.7564336)
(ZeroAccess,0.922786667)
};
        \addplot[fill=green]
            coordinates {
(Cridex,0.583)
(Harebot,0.789)
(Security Shield,0.662)
(Zbot,0.821)
(ZeroAccess,0.945)
};
        \addplot[fill=blue]
            coordinates {
(Cridex,0.619)
(Harebot,0.789)
(Security Shield,0.694)
(Zbot,0.812)
(ZeroAccess,0.945)
};
        \legend{Average HMM,Random restarts,Boosted HMMs}
    \end{axis}
\end{tikzpicture}
\caption{Initial experiments\label{bar:init_AUC}} 
\end{figure}
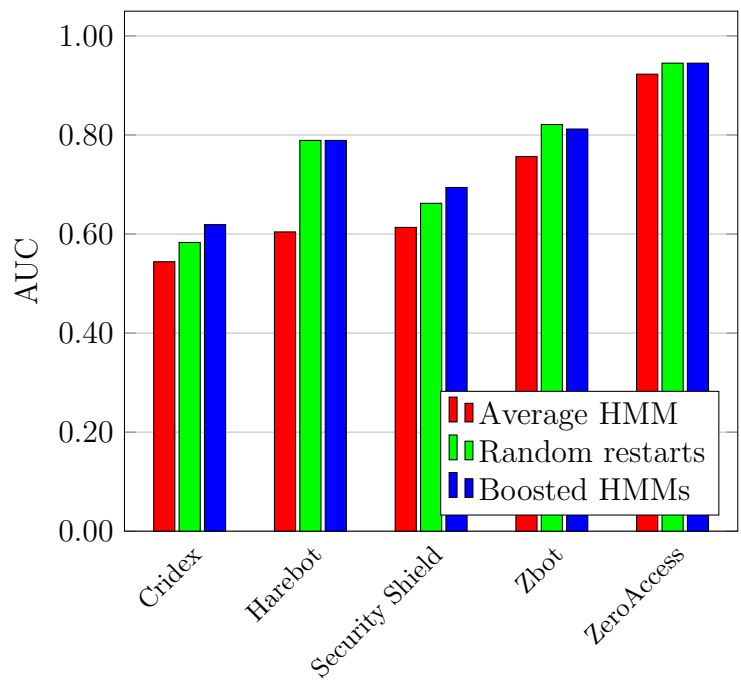

For our initial set of experiments, we trained models on each of the malware
families listed in Table~\ref{tab:BaseExpOpcodes}
and computed the AUC statistic.
In each case, we obtained results based on HMMs with~1000 random restarts, and 
also applied AdaBoost to classifiers based on these same~1000 HMM models.
The results for all of these experiments are summarized in the form of a bar graph
in Figure~\ref{bar:init_AUC}, where the ``average HMM'' is the 
average model over the~1000 HMMs. Thus, if we trained a single HMM,
we would expect to obtain the results given by the average HMM case.

From Figure~\ref{bar:init_AUC} we see multiple random restarts generally yields 
a significantly stronger model, and hence would, typically, likely be well worth the additional
(one time) work during the training phase. However, in most cases,
boosting offered minimal improvement over random restarts.
Specifically, boosting had little effect on the results for
Harebot, Zbot, and ZeroAccess. On the other hand, 
for Cridex and Security Shield, boosting does
provide some measurable improvement. Perhaps not surprisingly, 
it appears that the cases where boosting has something to offer 
are those where the non-boosted models are the weakest. 

It is also worth noting that boosting can significantly increase the work factor at
the scoring phase, since multiple HMMs are used in the
boosted classifier. This additional work depends on the number of HMM classifiers
used, and this can vary, as we select the optimal boosted classifiers
(i.e., the intermediate boosted model that
achieves the best results). For random restarts, we simply select the best model,
and hence the scoring phase is no more costly than a single HMM.
This points to an inherent advantage of the random restart case, and
hence we would generally only select the boosted model 
in cases where the improvement is significant. As noted above,
only Cridex and Security Shield offer any improvement
due to boosting, and the improvement in both cases is modest.
Therefore, we would argue that random restarts is the better choice
in all cases considered in this section.

In subsequent experiments we compare boosting 
and random restarts in more challenging classification problems.
Specifically, we consider the situation where the malware code is morphed,
for the purpose of making detection more difficult. Then we consider the so-called 
cold start problem, where the training data is limited.

\subsection{Morphing Experiments}\label{sec:morphing}

For our next set of experiments, we simulate code morphing, that is, 
we simulate the case where the malware
writer modifies the code in an attempt to make detection more difficult. 
Recall that we are considering detection based on differences in statistical properties 
of malware and benign opcode sequences. Therefore, we simulate the code morphing 
process by inserting opcode sequences extracted from benign
samples into malware samples. This should have the desired effect of 
making the malware samples statistically more similar to the benign samples, 
and hence make them far more difficult to distinguish from benign.

We consider three morphing cases. First, we insert benign code equivalent to~10\%\ of
the code, then we apply~50\%\ morphing and, finally, we use~100\%\ morphing. Note that,
for example, in the~100\%\ morphing case, the size of the opcode sequence doubles.
Our experimental results for these three cases
are summarized in Figure~\ref{bar:morph_AUC}.


\begin{figure}[!htbp]
\centering
\begin{tabular}{cc}
\begin{tikzpicture}[scale=0.675, every node/.style={scale=1.0}]
    \begin{axis}[
        width  = 0.6*\textwidth,
        height = 8cm,
        major x tick style = transparent,
        ybar=4*\pgflinewidth,
        bar width=8pt,
        ymajorgrids = true,
        ylabel = {AUC},
        symbolic x coords={Cridex, Harebot, Security Shield, Zbot, ZeroAccess},
	y tick label style={
    	/pgf/number format/.cd,
   	fixed,
   	fixed zerofill,
    	precision=2},
        xtick = data,
        x tick label style={rotate=45,anchor=north east, inner sep=0mm},
        scaled y ticks = false,
        enlarge x limits=0.135,
        ymin=0,
        ymax=1.05,
        legend cell align=left,
        legend pos=south east,
    ]
        \addplot[fill=red]
            coordinates {
(Cridex,0.57725)
(Harebot,0.52925)
(Security Shield,0.575)
(Zbot,0.756)
(ZeroAccess,0.857)
};
        \addplot[fill=green]
            coordinates {
(Cridex,0.636)
(Harebot,0.535)
(Security Shield,0.694)
(Zbot,0.755)
(ZeroAccess,0.869)
};
        \addplot[fill=blue]
            coordinates {
(Cridex,0.681)
(Harebot,0.556)
(Security Shield,0.712)
(Zbot,0.767)
(ZeroAccess,0.869)
};
        \legend{Average HMM,Random restarts,Boosted HMMs}
    \end{axis}
\end{tikzpicture}
&
\begin{tikzpicture}[scale=0.675, every node/.style={scale=1.0}]
    \begin{axis}[
        width  = 0.6*\textwidth,
        height = 8cm,
        major x tick style = transparent,
        ybar=4*\pgflinewidth,
        bar width=8pt,
        ymajorgrids = true,
        ylabel = {AUC},
        symbolic x coords={Cridex, Harebot, Security Shield, Zbot, ZeroAccess},
	y tick label style={
    	/pgf/number format/.cd,
   	fixed,
   	fixed zerofill,
    	precision=2},
        xtick = data,
        x tick label style={rotate=45,anchor=north east, inner sep=0mm},
        scaled y ticks = false,
        enlarge x limits=0.135,
        ymin=0,
        ymax=1.05,
        legend cell align=left,
        legend pos=south east,
    ]
        \addplot[fill=red]
            coordinates {
(Cridex,0.5775)
(Harebot,0.57925)
(Security Shield,0.689)
(Zbot,0.5275)
(ZeroAccess,0.646)
};
        \addplot[fill=green]
            coordinates {
(Cridex,0.685)
(Harebot,0.63)
(Security Shield,0.689)
(Zbot,0.654)
(ZeroAccess,0.646)
};
        \addplot[fill=blue]
            coordinates {
(Cridex,0.685)
(Harebot,0.667)
(Security Shield,0.689)
(Zbot,0.654)
(ZeroAccess,0.751)
};
        \legend{Average HMM,Random restarts,Boosted HMMs}
    \end{axis}
\end{tikzpicture}
\\
(a) Morphing at~10\%
&
(b) Morphing at~50\%
\\[2ex]
\begin{tikzpicture}[scale=0.675, every node/.style={scale=1.0}]
    \begin{axis}[
        width  = 0.6*\textwidth,
        height = 8cm,
        major x tick style = transparent,
        ybar=4*\pgflinewidth,
        bar width=8pt,
        ymajorgrids = true,
        ylabel = {AUC},
        symbolic x coords={Cridex, Harebot, Security Shield, Zbot, ZeroAccess},
	y tick label style={
    	/pgf/number format/.cd,
   	fixed,
   	fixed zerofill,
    	precision=2},
        xtick = data,
        x tick label style={rotate=45,anchor=north east, inner sep=0mm},
        scaled y ticks = false,
        enlarge x limits=0.135,
        ymin=0,
        ymax=1.05,
        legend cell align=left,
        legend pos=south east,
    ]
        \addplot[fill=red]
            coordinates {
(Cridex,0.63275)
(Harebot,0.597)
(Security Shield,0.68875)
(Zbot,0.6105)
(ZeroAccess,0.831)
};
        \addplot[fill=green]
            coordinates {
(Cridex,0.675)
(Harebot,0.661)
(Security Shield,0.71)
(Zbot,0.636)
(ZeroAccess,0.849)
};
        \addplot[fill=blue]
            coordinates {
(Cridex,0.675)
(Harebot,0.703)
(Security Shield,0.71)
(Zbot,0.64)
(ZeroAccess,0.892)
};
        \legend{Average HMM,Random restarts,Boosted HMMs}
    \end{axis}
\end{tikzpicture}
\\
(c) Morphing at~100\%
\end{tabular}
\caption{Morphing experiments\label{bar:morph_AUC}} 
\end{figure}

As with the previous experiments, the differences between the average HMM
and the best of the random restarts model is generally significant.
On the other hand, the differences between the random restarts and the boosted model are not
large in most cases, but there are some cases where the differences are significant.
For example, Cridex at~10\%\ morphing and ZeroAccess at~50\%\ morphing
both show substantial improvement for the boosted models, as compared to
random restarts. Interestingly, there does not seem to be a clear trend 
as to when boosting is likely to offer more than a marginal improvement.
Again, we would likely choose random restarts in all of these cases, as
the improvement provided by boosting is insufficient to justify the
additional work required when scoring via boosted models.

\subsection{Cold Start Problem}\label{sec:cold_start_problem}

In machine learning, the cold start problem deals with the case where the training data is severely 
limited. This is of practical concern when attempting to detect malware, as initially we might only have
a small number of samples available for training. In such cases, it would be important to know
how much data is needed before reliable models can be generated. And, the
cold start problem is particularly relevant to the research here, as we are considering
methods to improve our classification results by training a large number of models.
Intuitively, these techniques are likely to be most needed in marginal cases,
and the cold start problem provides just such a case.

For our cold start experiments,
we varied the training data size from~5 to~25 samples, in increments of~5. For the 
families with a large number of samples available
(Zbot and ZeroAccess), we tested models on~200 malware samples in each case, 
so that the malware and benign sets are more in balance. For the remaining families 
(Cridex, Harebot, and Security Shield), which have few samples,
we used all of the non-training sample for testing.
Recall that we have~107 samples in our benign set. 

The results of our cold start experiments are summarized
in Figures~\ref{fig:cs_AUC_bar_families} and~\ref{fig:cs_accuracy_bar_families}, 
in terms of AUC and accuracy, respectively. 
Note that each bar graph in both of these figures
includes results for the typical HMM case (the red bar), as well as the result for 
the case where we select the best model from the random restarts (the green bar), 
and the case where we apply AdaBoost to the models (blue bar).
All of these results are based on~1000 models, each generated
with a random initialization for the HMM training.

\begin{figure}[!htbp]
\centering
\begin{tabular}{cc}
\begin{tikzpicture}[scale=0.675, every node/.style={scale=0.675}]
    \begin{axis}[
        width  = 8cm,
        height = 8cm,
        major x tick style = transparent,
        ybar=4*\pgflinewidth,
        bar width=8.25pt,
        ymajorgrids = true,
        ylabel = {AUC},
        symbolic x coords={5 samples,10 samples,15 samples,20 samples,25 samples},
	y tick label style={
    	/pgf/number format/.cd,
   	fixed,
   	fixed zerofill,
    	precision=1},
        xtick = data,
        x tick label style={rotate=45,anchor=north east, inner sep=0mm},
        scaled y ticks = false,
        enlarge x limits=0.125,
        ymin=0,
        ymax=1.0,
        ytick = {0.0,0.2,0.4,0.6,0.8,1.0},
        legend cell align=left,
        legend pos=south east,
    ]
        \addplot[fill=red]
            coordinates {
(5 samples,0.530806837)
(10 samples,0.627300334)
(15 samples,0.573113604)
(20 samples,0.672491674)
(25 samples,0.705529306)
};
        \addplot[fill=green]
            coordinates {
(5 samples,0.628742)
(10 samples,0.695678)
(15 samples,0.638365)
(20 samples,0.786431)
(25 samples,0.797063)
};
        \addplot[fill=blue]
            coordinates {
(5 samples,0.741027)
(10 samples,0.820897)
(15 samples,0.798194)
(20 samples,0.81819)
(25 samples,0.88642)
};
        \legend{Average HMM, Random restarts, Boosted HMMs}
    \end{axis}
\end{tikzpicture}
&
\begin{tikzpicture}[scale=0.675, every node/.style={scale=0.675}]
    \begin{axis}[
        width  = 8cm,
        height = 8cm,
        major x tick style = transparent,
        ybar=4*\pgflinewidth,
        bar width=8.25pt,
        ymajorgrids = true,
        ylabel = {AUC},
        symbolic x coords={5 samples,10 samples,15 samples,20 samples,25 samples},
	y tick label style={
    	/pgf/number format/.cd,
   	fixed,
   	fixed zerofill,
    	precision=1},
        xtick = data,
        x tick label style={rotate=45,anchor=north east, inner sep=0mm},
        scaled y ticks = false,
        enlarge x limits=0.125,
        ymin=0,
        ymax=1.0,
        ytick = {0.0,0.2,0.4,0.6,0.8,1.0},
        legend cell align=left,
        legend pos=south east,
    ]
        \addplot[fill=red]
            coordinates {
(5 samples,0.546003516)
(10 samples,0.582051984)
(15 samples,0.559338029)
(20 samples,0.586687326)
(25 samples,0.559858583)
};
        \addplot[fill=green]
            coordinates {
(5 samples,0.586059)
(10 samples,0.617257)
(15 samples,0.61879)
(20 samples,0.691022)
(25 samples,0.63785)
};
        \addplot[fill=blue]
            coordinates {
(5 samples,0.719724)
(10 samples,0.791676)
(15 samples,0.775455)
(20 samples,0.829085)
(25 samples,0.79773)
};
        \legend{Average HMM, Random restarts, Boosted HMMs}
    \end{axis}
\end{tikzpicture}
\\
(a) Cridex & (b) Harebot
\\[2ex]
\begin{tikzpicture}[scale=0.675, every node/.style={scale=0.675}]
    \begin{axis}[
        width  = 8cm,
        height = 8cm,
        major x tick style = transparent,
        ybar=4*\pgflinewidth,
        bar width=8.25pt,
        ymajorgrids = true,
        ylabel = {AUC},
        symbolic x coords={5 samples,10 samples,15 samples,20 samples,25 samples},
	y tick label style={
    	/pgf/number format/.cd,
   	fixed,
   	fixed zerofill,
    	precision=1},
        xtick = data,
        x tick label style={rotate=45,anchor=north east, inner sep=0mm},
        scaled y ticks = false,
        enlarge x limits=0.125,
        ymin=0,
        ymax=1.0,
        ytick = {0.0,0.2,0.4,0.6,0.8,1.0},
        legend cell align=left,
        legend pos=south east,
    ]
        \addplot[fill=red]
            coordinates {
(5 samples,0.580149429)
(10 samples,0.594950308)
(15 samples,0.609062246)
(20 samples,0.618282002)
(25 samples,0.538770062)
};
        \addplot[fill=green]
            coordinates {
(5 samples,0.602716)
(10 samples,0.614681)
(15 samples,0.687459)
(20 samples,0.802263)
(25 samples,0.739734)
};
        \addplot[fill=blue]
            coordinates {
(5 samples,0.715747)
(10 samples,0.65625)
(15 samples,0.77331)
(20 samples,0.772012)
(25 samples,0.749166)
};
        \legend{Average HMM, Random restarts, Boosted HMMs}
    \end{axis}
\end{tikzpicture}
&
\begin{tikzpicture}[scale=0.675, every node/.style={scale=0.675}]
    \begin{axis}[
        width  = 8cm,
        height = 8cm,
        major x tick style = transparent,
        ybar=4*\pgflinewidth,
        bar width=8.25pt,
        ymajorgrids = true,
        ylabel = {AUC},
        symbolic x coords={5 samples,10 samples,15 samples,20 samples,25 samples},
	y tick label style={
    	/pgf/number format/.cd,
   	fixed,
   	fixed zerofill,
    	precision=1},
        xtick = data,
        x tick label style={rotate=45,anchor=north east, inner sep=0mm},
        scaled y ticks = false,
        enlarge x limits=0.125,
        ymin=0,
        ymax=1.0,
        ytick = {0.0,0.2,0.4,0.6,0.8,1.0},
        legend cell align=left,
        legend pos=south east,
    ]
        \addplot[fill=red]
            coordinates {
(5 samples,0.76865552)
(10 samples,0.85094671)
(15 samples,0.570559483)
(20 samples,0.657038704)
(25 samples,0.623924617)
};
        \addplot[fill=green]
            coordinates {
(5 samples,0.91243)
(10 samples,0.888598)
(15 samples,0.84271)
(20 samples,0.72257)
(25 samples,0.781776)
};
        \addplot[fill=blue]
            coordinates {
(5 samples,0.943621)
(10 samples,0.919486)
(15 samples,0.879136)
(20 samples,0.785187)
(25 samples,0.811168)
};
        \legend{Average HMM, Random restarts, Boosted HMMs}
    \end{axis}
\end{tikzpicture}
\\
(c) Security Shield & (d) Zbot
\\[2ex]
\begin{tikzpicture}[scale=0.675, every node/.style={scale=0.675}]
    \begin{axis}[
       width  = 8cm,
       height = 8cm,
        major x tick style = transparent,
        ybar=4*\pgflinewidth,
        bar width=8.25pt,
        ymajorgrids = true,
        ylabel = {AUC},
        symbolic x coords={5 samples,10 samples,15 samples,20 samples,25 samples},
	y tick label style={
    	/pgf/number format/.cd,
   	fixed,
   	fixed zerofill,
    	precision=1},
        xtick = data,
        x tick label style={rotate=45,anchor=north east, inner sep=0mm},
        scaled y ticks = false,
        enlarge x limits=0.125,
        ymin=0,
        ymax=1.0,
        ytick = {0.0,0.2,0.4,0.6,0.8,1.0},
        legend cell align=left,
        legend pos=south east,
    ]
        \addplot[fill=red]
            coordinates {
(5 samples,0.954084793)
(10 samples,0.912126308)
(15 samples,0.957691185)
(20 samples,0.943398576)
(25 samples,0.896308)
};
        \addplot[fill=green]
            coordinates {
(5 samples,0.989626)
(10 samples,0.933458)
(15 samples,0.990561)
(20 samples,0.976402)
(25 samples,0.98271)
};
        \addplot[fill=blue]
            coordinates {
(5 samples,0.976822)
(10 samples,0.864696)
(15 samples,0.976075)
(20 samples,0.952897)
(25 samples,0.975421)
};
        \legend{Average HMM, Random restarts, Boosted HMMs}
    \end{axis}
\end{tikzpicture}
\\
(e) ZeroAccess
\end{tabular}
\caption{Cold start AUC results by family}\label{fig:cs_AUC_bar_families}
\end{figure}
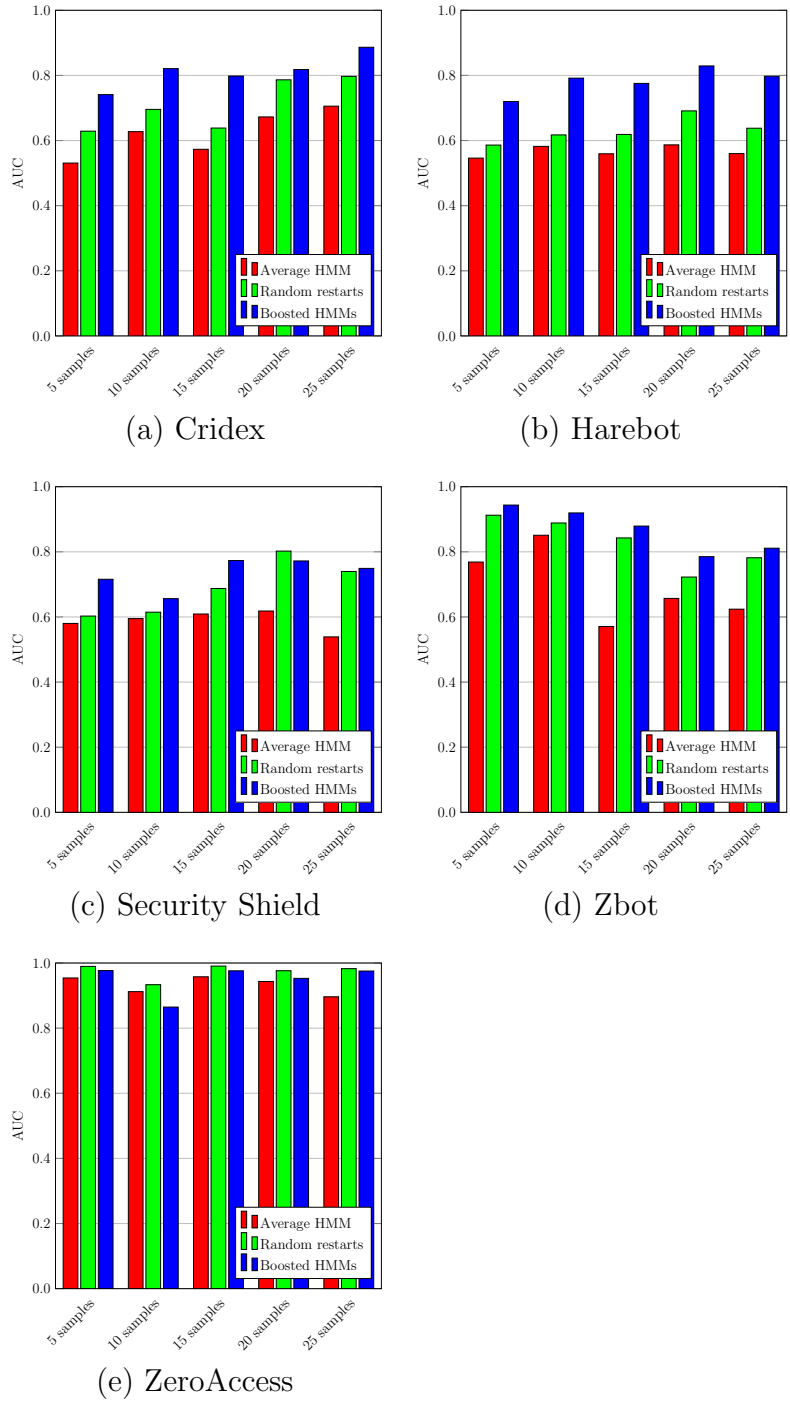

\begin{figure}[!htbp]
\centering
\begin{tabular}{cc}
\begin{tikzpicture}[scale=0.675, every node/.style={scale=0.9}]
    \begin{axis}[
        width  = 8cm,
        height = 8cm,
        major x tick style = transparent,
        ybar=4*\pgflinewidth,
        bar width=8.25pt,
        ymajorgrids = true,
        ylabel = {Accuracy},
        symbolic x coords={5 samples,10 samples,15 samples,20 samples,25 samples},
	y tick label style={
    	/pgf/number format/.cd,
   	fixed,
   	fixed zerofill,
    	precision=1},
        xtick = data,
        x tick label style={rotate=45,anchor=north east, inner sep=0mm},
        scaled y ticks = false,
        enlarge x limits=0.125,
        ymin=0,
        ymax=1.0,
        ytick = {0.0,0.2,0.4,0.6,0.8,1.0},
        legend cell align=left,
        legend pos=south east,
    ]
        \addplot[fill=red]
            coordinates {
(5 samples,0.706767052)
(10 samples,0.785110898)
(15 samples,0.749705029)
(20 samples,0.80270196)
(25 samples,0.834769387)
};
        \addplot[fill=green]
            coordinates {
(5 samples,0.772727)
(10 samples,0.818713)
(15 samples,0.801205)
(20 samples,0.850932)
(25 samples,0.891026)
};
        \addplot[fill=blue]
            coordinates {
(5 samples,0.778409)
(10 samples,0.853801)
(15 samples,0.837349)
(20 samples,0.857143)
(25 samples,0.916667)
};
        \legend{Average HMM, Random restarts, Boosted HMMs}
    \end{axis}
\end{tikzpicture}
&
\begin{tikzpicture}[scale=0.675, every node/.style={scale=0.9}]
    \begin{axis}[
        width  = 8cm,
        height = 8cm,
        major x tick style = transparent,
        ybar=4*\pgflinewidth,
        bar width=8.25pt,
        ymajorgrids = true,
        ylabel = {Accuracy},
        symbolic x coords={5 samples,10 samples,15 samples,20 samples,25 samples},
	y tick label style={
    	/pgf/number format/.cd,
   	fixed,
   	fixed zerofill,
    	precision=1},
        xtick = data,
        x tick label style={rotate=45,anchor=north east, inner sep=0mm},
        scaled y ticks = false,
        enlarge x limits=0.125,
        ymin=0,
        ymax=1.0,
        ytick = {0.0,0.2,0.4,0.6,0.8,1.0},
        legend cell align=left,
        legend pos=south east,
    ]
        \addplot[fill=red]
            coordinates {
(5 samples,0.752251527)
(10 samples,0.818726534)
(15 samples,0.797027495)
(20 samples,0.82608588)
(25 samples,0.823881438)
};
        \addplot[fill=green]
            coordinates {
(5 samples,0.787097)
(10 samples,0.853333)
(15 samples,0.855172)
(20 samples,0.871429)
(25 samples,0.874074)
};
        \addplot[fill=blue]
            coordinates {
(5 samples,0.8)
(10 samples,0.86)
(15 samples,0.862069)
(20 samples,0.885714)
(25 samples,0.881481)
};
        \legend{Average HMM, Random restarts, Boosted HMMs}
    \end{axis}
\end{tikzpicture}
\\
(a) Cridex & (b) Harebot
\\[1ex]
\begin{tikzpicture}[scale=0.675, every node/.style={scale=0.9}]
    \begin{axis}[
        width  = 8cm,
        height = 8cm,
        major x tick style = transparent,
        ybar=4*\pgflinewidth,
        bar width=8.25pt,
        ymajorgrids = true,
        ylabel = {Accuracy},
        symbolic x coords={5 samples,10 samples,15 samples,20 samples,25 samples},
	y tick label style={
    	/pgf/number format/.cd,
   	fixed,
   	fixed zerofill,
    	precision=1},
        xtick = data,
        x tick label style={rotate=45,anchor=north east, inner sep=0mm},
        scaled y ticks = false,
        enlarge x limits=0.125,
        ymin=0,
        ymax=1.0,
        ytick = {0.0,0.2,0.4,0.6,0.8,1.0},
        legend cell align=left,
        legend pos=south east,
    ]
        \addplot[fill=red]
            coordinates {
(5 samples,0.7876875)
(10 samples,0.782245472)
(15 samples,0.781793136)
(20 samples,0.856103249)
(25 samples,0.852078613)
};
        \addplot[fill=green]
            coordinates {
(5 samples,0.79375)
(10 samples,0.787097)
(15 samples,0.84)
(20 samples,0.868966)
(25 samples,0.864286)
};
        \addplot[fill=blue]
            coordinates {
(5 samples,0.80625)
(10 samples,0.787097)
(15 samples,0.84)
(20 samples,0.875862)
(25 samples,0.871429)
};
        \legend{Average HMM, Random restarts, Boosted HMMs}
    \end{axis}
\end{tikzpicture}
&
\begin{tikzpicture}[scale=0.675, every node/.style={scale=0.9}]
    \begin{axis}[
        width  = 8cm,
        height = 8cm,
        major x tick style = transparent,
        ybar=4*\pgflinewidth,
        bar width=8.25pt,
        ymajorgrids = true,
        ylabel = {Accuracy},
        symbolic x coords={5 samples,10 samples,15 samples,20 samples,25 samples},
	y tick label style={
    	/pgf/number format/.cd,
   	fixed,
   	fixed zerofill,
    	precision=1},
        xtick = data,
        x tick label style={rotate=45,anchor=north east, inner sep=0mm},
        scaled y ticks = false,
        enlarge x limits=0.125,
        ymin=0,
        ymax=1.0,
        ytick = {0.0,0.2,0.4,0.6,0.8,1.0},
        legend cell align=left,
        legend pos=south east,
    ]
        \addplot[fill=red]
            coordinates {
(5 samples,0.738029432)
(10 samples,0.812925232)
(15 samples,0.66073304)
(20 samples,0.695153056)
(25 samples,0.663026126)
};
        \addplot[fill=green]
            coordinates {
(5 samples,0.879479)
(10 samples,0.846906)
(15 samples,0.807818)
(20 samples,0.703583)
(25 samples,0.736156)
};
        \addplot[fill=blue]
            coordinates {
(5 samples,0.912052)
(10 samples,0.876221)
(15 samples,0.85342)
(20 samples,0.726384)
(25 samples,0.771987)
};
        \legend{Average HMM, Random restarts, Boosted HMMs}
    \end{axis}
\end{tikzpicture}
\\
(c) Security Shield & (d) Zbot
\\[1ex]
\begin{tikzpicture}[scale=0.675, every node/.style={scale=0.9}]
    \begin{axis}[
       width  = 8cm,
       height = 8cm,
        major x tick style = transparent,
        ybar=4*\pgflinewidth,
        bar width=8.25pt,
        ymajorgrids = true,
        ylabel = {Accuracy},
        symbolic x coords={5 samples,10 samples,15 samples,20 samples,25 samples},
	y tick label style={
    	/pgf/number format/.cd,
   	fixed,
   	fixed zerofill,
    	precision=1},
        xtick = data,
        x tick label style={rotate=45,anchor=north east, inner sep=0mm},
        scaled y ticks = false,
        enlarge x limits=0.125,
        ymin=0,
        ymax=1.0,
        ytick = {0.0,0.2,0.4,0.6,0.8,1.0},
        legend cell align=left,
        legend pos=south east,
    ]
        \addplot[fill=red]
            coordinates {
(5 samples,0.910713043)
(10 samples,0.85319549)
(15 samples,0.916012791)
(20 samples,0.914195566)
(25 samples,0.914915273)
};
        \addplot[fill=green]
            coordinates {
(5 samples,0.973941)
(10 samples,0.879479)
(15 samples,0.970684)
(20 samples,0.947883)
(25 samples,0.970684)
};
        \addplot[fill=blue]
            coordinates {
(5 samples,0.973941)
(10 samples,0.879479)
(15 samples,0.973941)
(20 samples,0.954397)
(25 samples,0.977199)
};
        \legend{Average HMM, Random restarts, Boosted HMMs}
    \end{axis}
\end{tikzpicture}
\\
(e) ZeroAccess
\end{tabular}
\vglue-0.075in
\caption{Cold start accuracy results by family}\label{fig:cs_accuracy_bar_families}
\end{figure}

The general trends for the AUC and accuracy results are similar, so
we discuss only the accuracy graphs in Figure~\ref{fig:cs_accuracy_bar_families};
similar comments hold for the the AUC graphs in Figure~\ref{fig:cs_AUC_bar_families}
For Cridex, Harebot, and Security Shield, we see a generally upward trend
in Figures~\ref{fig:cs_accuracy_bar_families}~(a), (b), and~(c), respectively, as the number
of training samples increases. For the ZeroAccess family in 
Figure~\ref{fig:cs_accuracy_bar_families}~(e), we see little change, indicating
that~5 samples is apparently sufficient to obtain essentially optimal results.
The Zbot family in Figure~\ref{fig:cs_accuracy_bar_families}~(d) is somewhat
anomalous---and somewhat surprising---as we see a downward trend in the accuracy
with an increase the number of training samples. 
The Zbot results in Figure~\ref{fig:cs_accuracy_bar_families}~(d) seem to indicate
that for this particular family, the models generated are unstable, 
in the sense that the models depend heavily on the specific samples selected
for training.

With respect to the three approaches considered in this section
(i.e., typical hidden Markov model, multiple random
restarts, and boosted HMMs), the results in Figure~\ref{fig:cs_accuracy_bar_families} 
show a significant advantage for multiple random restarts over the average
HMM in almost every case. The advantage of boosting over multiple
random restarts is certainly less pronounced, but is sginficant 
in some cases---and this advantage is generally greatest
for the cases where classification is the most challenging. These
results indicate that  boosting is likely to only be worthwhile
in extremely challenging cases. Taking into account the additional work
required for scoring using a boosted model only serves to further
emphasize this point.

%

\section{Conclusion and Future Work}\label{chap:cl_fw}

With the increasing threat of malware, it is critically important to have the most efficient 
and effective malware detection techniques possible. In this paper, we have explored 
improved classifier methods based on hidden Markov models, 
using both random restarts and boosting. We found that training multiple HMMs with
different initial values will generally yield significant improvement over 
generating a single HMM. Multiple random restarts adds work in the training
phase, but the scoring phase is no more costly, since we use only a single HMM.
Since training is one-time work, in many cases it would be reasonable to 
train a large number of models using random restarts.

For our boosting experiments, we used AdaBoost, which is 
straightforward and inexpensive in the training phase. The improvement offered by boosting over
multiple random restarts was, in general, surprisingly small, but in some of the most challenging cases,
boosting did offer a significant improvement. However, boosted classifiers 
are significantly more costly in the scoring phase, as a large number of 
models are typically used in the final boosted classifier. Furthermore, 
boosting is not particularly robust, in the sense
that errors in the training data tend to grow when training the
boosted classifier~\cite{InfoSecAdaBoost}. 
Consequently, we would likely only use boosting in situations 
where the improvement is significant, as compared to the non-boosted case.

For future work, it would be worthwhile to consider boosting for malware detection,
based on machine learning models other than HMMs. It would also be interesting to
consider data contamination attacks, which would tend to have a larger negative impact on 
boosted models than on non-boosted models. Under these and other scenarios, it would
be valuable, albeit challenging, to determine conditions under which
boosting is likely to yield a classifier that is sufficiently strong to justify
the additional cost and risk.

\bibliographystyle{abbrv}

\bibliography{references}

\end{document}